# Gamma Rays from Compact Objects

*Ludwig Biermann Award Lecture* [1]

Karl Mannheim

*Universitäts-Sternwarte, Geismarlandstrasse 11, D-37803 Göttingen, Germany;*
*kmannhe@uni-sw.gwdg.de*

Οὕτω δὴ στάσιν μὲν ἐν ὁμαλότητι, κίνησιν δὲ εἰς ἀνωμαλότητα ἀεὶ τιθῶμεν. αἰτία δὲ ἀνισότης αὖ τῆς ἀνωμάλου φύσεως [2].     Platon : Timaios

**Abstract**

Pulsars, Gamma Ray Bursts and gamma ray emitting Active Galactic Nuclei represent the most luminous compact gamma ray sources in the Universe. Based on their apparent luminosities and non-thermal spectra, a direct association of the gamma ray sources with objects of the size of a neutron star or an accreting black hole appears questionable at a first glance: Compact radiation fields are optically thick to pair creation. The paradox can be resolved, if the gamma ray emitting plasma is in relativistic bulk motion with respect to the observer. The relativistically expanding plasma dissipates a significant fraction of its bulk kinetic energy into gamma rays, presumably through particle acceleration at collisionless shocks. Details of the conversion of bulk kinetic energy, carried by baryons, into radiation, emitted by leptons, are highly controversial. Results are important for the problem of the origin of the highest energy cosmic rays and the composition of the plasma emerging from the ergosphere of compact objects. Measurements of ultra-high energy neutrinos can be used to discriminate between baryon-induced electromagnetic cascade or lepton-induced inverse-Compton models.

Gamma rays from remote quasars at high redshifts may be used to probe the density of diffuse intergalactic infrared-to-ultraviolet photons from the era of galaxy formation. An exponential turnover in the gamma ray spectra marks the onset of pair creation by photon-photon interactions between the quasar gamma rays and diffuse low energy photons along the line of sight. If the background density can be inferred from direct observations, one could determine the Hubble constant at unprecedented distances where distortions of the cosmological expansion due to mass flows accelerated by local mass concentrations can be neglected.

---

[1] submitted to Rev.Mod.Astr.

[2] "Thus we conclude to attribute the steady-state to homology, kinetics however to anomaly. Anisotropy, on the other hand, is the reason for the anomalous nature."

# 1 Introduction

Ludwig Biermann's (1951) discovery of the interaction between the corpuscular radiation from the sun and the ions in the tails of comets has forcefully demonstrated the importance of *dynamical* processes in the solar system. Satellite experiments demonstrate that the interplanetary medium experiences a wealth of phenomena leading to particle acceleration and non-thermal emission. Since the solar wind is rather slow, these non-thermal processes are only of local importance. However, the stellar wind of rapidly rotating stars and the ejecta of supernovae reach much higher velocities and, consequently, their non-thermal emission is much stronger. Radio synchrotron emission from the expanding shell of a supernova indicates the acceleration of particles to relativistic energies at the blast wave. Supernova remnants are the most likely origin of the cosmic ray particles below $\sim 10^{15}$ eV (Axford 1981) or, possibly, to even higher energies (Völk & Biermann 1988). The presence of a frozen magnetic field in the highly conducting plasma introduces anisotropies in the phase-space distribution of particles driving them away from thermodynamic equilibrium. Energetic particles in the low density plasma cannot maintain detailed balance by rare Coulomb collisions. The gyrating particles are forced to phase-space diffusion through a resonant coupling with the collective degrees of freedom of the magnetic field fluctuations which are correlated by turbulent cascades. Today we know that stellar winds are a common phenomenon, and often intimately linked with the presence of rotation and magnetic fields. Parker (1958) has realized that the lever arm provided by the corotation of the stellar wind below the Alfvén point along each streamline acts as a breaking torque extracting rotational energy and angular momentum from the star. It is therefore not surprising that rapidly rotating magnetized neutron stars and accretion disks around black holes produce powerful magnetized winds. Rotating disk winds can self-collimate attaining a bipolar topology as found in radio jets emerging from many AGNs and some galactic black hole candidates (Camenzind 1990).

In this Lecture, I will introduce the audience to various phenomena of gamma ray astrophysics and their possible explanation as the consequence of particle acceleration in the plasma flows emerging from compact objects, *viz.* Active Galactic Nuclei (AGNs), Gamma Ray Bursts (GRBs) and pulsars. Emphasis of the Lecture is on AGN jets. Experimental progress in this field of astrophysics has been very rapid in recent years. Astronomical objects can now be observed from radio waves to TeV gamma rays using a diversity of telescopes, from radio antennas to Čerenkov mirrors and scintillator arrays providing sensitivity for 20 orders of magnitude in photon energy. More than a thousand GRBs (BATSE 3B catalog) and a surprisingly large number of AGNs ($\sim 50$) have been detected by the Compton Gamma Ray Observatory (Fichtel et al. 1994, Thompson et al. 1995). Using ground-based Čerenkov telescopes, steady TeV ($10^{12}$ eV) emission has been discovered from two pulsars, the Crab nebula and PSR 1706-044 (Weekes 1994), and from two AGNs, Mrk401 (Punch et al. 1992) and Mrk501 (Quinn et al. 1996). The gamma ray emitting AGNs belong to the class of blazars characterized by drastic variability at all wavelengths, a high degree of linear polarization and a broad-band continuum spectrum. Morphologically, the sources show compact radio cores and often jets. A common spectral characteristic of blazars, GRBs and the pulsed phase of pulsars (especially old pulsars) is that the energy flux is typically dominated by gamma rays (Montigny et al. 1995). Models of the radiation processes responsible for the ultraluminous gamma ray emission allow for a diagnosis of the plasma conditions in the vicinity of the compact objects powering the sources and along the line of sight.

Gamma ray astrophysics brings together astronomy and physics on several important topics. Stable supersymmetric particles have escaped their detection in accelerator experiments because of their large masses, but they could be detected through gravitational interaction on astronomical scales. Their existence could solve the long-standing mystery of the missing mass inferred from flat galactic rotation curves or galaxy cluster dynamics. Direct evidence could be obtained by measuring high-energy neutrinos from dark matter particle–antiparticle annihilation in astronomical bodies where dark matter accumulates over the Hubble time (Kamionkowski 1992, Gaisser et al. 1994). Astrophysical gamma rays indicate the possible existence of a diffuse flux of high-energy neutrinos which would constitute a conservative background for searches of such 'new physics'. Astronomy provides many more rather challenging phenomena than just dark matter. There is mounting evidence for black holes, with the best evidence to date for the supermassive black hole in NGC4258 inferred from the Keplerian rotation of water-maser emitting clouds (Miyoshi et al. 1995). It is an unsolved problem in physics whether the existence of the event horizon, which develops during a gravitational collaps, implies the existence of a space-time singularity in its interior. By Penrose's theorem, a singularity must exist if the general theory of relativity is true. Only the yet undiscovered laws of quantum-gravity could forbid the formation of a singularity which we render unphysical based on our scientific belief. In a sense, the existence of black holes is telling us that a theory of quantum-gravity must exist. Astronomical observations can bring us very close to the event horizon in *accreting* black hole systems where matter is dissipatively sinking into the potential well of the black hole. As Ludwig Biermann's corpuscular radiation from the sun, these black hole systems eject high speed particles along their angular momentum axes. The nature of these particles is still poorly understood, and in this Lecture I will present diagnostic strategies based on gamma ray and neutrino astronomy. Another exciting problem comes from cosmic ray astrophysics where trans-Greisen particles with energies exceeding $5 \times 10^{19}$ eV have been discovered recently. Based on the shower properties, the particles are most probably protons (Dedenko et al. 1995). Due to photo-production of pions in the 3K-background radiation field, cosmic ray protons above the Greisen-energy can not propagate more than ∼100 Mpc (Greisen 1966, Zatsepin & Kuz´min 1966). Where do the protons come from? They could be related to topological defects predicted by inflationary cosmology due to a phase transition in the early Universe (Sigl et al. 1994). However, new physics is not necessarily required by these exotic events, if protons are shock-accelerated in the jets of radio galaxies (Biermann & Strittmatter 1987).

The plan of the Lecture is as follows: In Sect.2 I will confront the audience with the striking paradox that the observed gamma ray point sources are too compact to emit gamma rays by introducing the important process of photon-photon pair creation. It is shown that relativistic bulk motion resolves this paradox. In Sect.3 I will outline a scenario of accreting supermassive black holes driving a relativistic jet as the central powerhouse in AGNs. Sect.4 will zoom-in on the proton-induced cascade mechanism for the gamma ray emission in AGNs. In addition, flux predictions for ultrahigh energy neutrinos are presented and compared with experimental sensitivities in Sect.5. A natural source for neutrinos with energies exceeding the electroweak phase transition energy would unlock the door to a new era of particle astrophysics. Finally, the notion of a gamma ray horizon due to intergalactic absorption of gamma rays is introduced in Sect.6. Measurements of gamma ray absorption may provide important clues on the cosmic distance scale and on galaxy formation. In addition, a strategy is developed to solve the halo vs. cosmological origin problem for GRBs.

## 2 The Paradox

Let us briefly look at the gamma ray luminosities and sizes of pulsars, GRBs and AGNs in Tab.1. The luminosities given in the first column correspond to the isotropic luminosity

$$L = 4\pi D^2 \int_{E_{\min}}^{E_{\max}} E \frac{dN}{dE} dE \quad (1)$$

where $D$ denotes the distance (for GRBs, see discussion after Eq.(13)), $E$ the observed photon energy in the gamma ray band between $E_{\min}$ and $E_{\max}$ for a photon number distribution $dN/dE = N_\circ (E/E_\circ)^{-(\alpha+1)}$. The sizes in the second column correspond to the neutron star radius $r_\circ \sim \sqrt{\hbar c/G m_{\rm p}^2} \hbar/m_{\rm p} c \sim$ 3 km or the Schwarzschild radius $r_\circ = r_{\rm S} = 2GM/c^2 = 3 \times 10^{15} m_{10}$ cm of an Eddington radiating black hole of mass $M = 10^{10} m_{10} M_\odot$, respectively. Column three shows the upper limit on the source size $r_{\rm var} < c\Delta t$ due to the requirement that the light travel time across the source is less than the observed variability time scale[3] $\Delta t$. The fourth column denotes the optical

---
[3]Incoherent emission is tacitly assumed here. If coherent emission is important, the causality constraint would be invalidated. However, the size of phase-space is $\propto E^{-3}$

depth to pair creation at a photon energy of $E = 2m_e c^2 \simeq 1$ MeV (Gould & Shréder 1966), i.e. the process

$$\gamma + \gamma \to e^+ + e^- \tag{2}$$

Pair creation can only occur, if the two interacting photons provide more than twice the electron rest mass in their center-of-momentum frame, that is if the target photon energy $\epsilon$ exceeds the threshold energy $\epsilon_{\rm th}$

$$E\epsilon_{\rm th}(1+z)^2(1-\cos\theta_{\rm s}) = 2(m_e c^2)^2 \tag{3}$$

in a collision at an angle $\theta_{\rm s}$. The $(1+z)^2$ factor takes into account cosmological or gravitational redshift. The interaction cross section is given by

$$\sigma_{\gamma\gamma} = \frac{3\sigma_{\rm T}}{16}(1-\beta^2)\left[2\beta(\beta^2-2) + (3-\beta^4)\ln\left(\frac{1+\beta}{1-\beta}\right)\right] \tag{4}$$

where $\beta = \sqrt{1 - 1/u^2}$ with $u^2 = \epsilon/\epsilon_{\rm th}$, and where $\sigma_{\rm T}$ denotes the Thomson cross section. For an average collision in an isotropic target radiation field ($\cos\theta_{\rm s} = 0$), the cross section Eq.(4) peaks at $\sim \sigma_{\rm T}/3$ when $E \sim 2(m_e c^2)^2/\epsilon$. The optical depth $\tau_{\gamma\gamma}$ for gamma rays of energy $E = 2m_e c^2$ is then given by

$$\tau_{\gamma\gamma,{\rm o}} \simeq n_\gamma(\geq m_e c^2)\frac{1}{3}\sigma_{\rm T} r \tag{5}$$

which is of sufficient accuracy for the qualitative comparison in Tab.1. An inverse power law distribution of target photons $n_\gamma(>\epsilon) = \int_\epsilon^\infty dn \propto \epsilon^{-\alpha}$ leads to the energy-dependent optical depth

$$\tau_{\gamma\gamma}(E) = \tau_{\gamma\gamma,{\rm o}}(E/2m_e c^2)^\alpha \quad , \tag{6}$$

since the threshold condition Eq.(3) requires that the target photon energy obeys $\epsilon_{\rm th} \propto 1/E$, so that $n_\gamma(\geq \epsilon_{\rm th}) \propto \epsilon_{\rm th}^{-\alpha} \propto E^\alpha$.

From Tab.1 it is evident that the optical depth for $E > 2m_e c^2$ exceeds unity *implying the paradoxical result that the objects cannot emit non-thermal gamma rays above MeV*.

PULSARs: For the estimate of the optical depth of the Crab pulsar magnetosphere, I have used the radius $r_{\rm o} = 10$ km instead of $c\Delta t$ which is more appropriate for the emission from the pulsar wind. The pulsar wind has a very low gamma ray optical depth and there is no paradox for the unpulsed emission from the nebula. However, for the pulsed component the paradox becomes striking considering the energy dependence in Eq.(6). Pulsed gamma rays have been observed, e.g. by EGRET at photon energies of 10 GeV yielding $\tau_{\gamma\gamma} \sim 100$ ($\alpha_{\rm x} = 0.5$). An obvious way out of the problem is to assume

---

making it very unlikely that coherent gamma ray processes occur in nature.

Table 1: Apparent pair creation optical depths for various gamma ray sources

| Type of Source | $L$ [erg/s] | $r_\mathrm{o}$ [cm] | $c\Delta t$ [cm] | $\tau_{\gamma\gamma,\mathrm{o}}$ | $\Gamma_\mathrm{o}$ |
|---|---|---|---|---|---|
| Pulsar (Crab[a]) | $10^{36}$ | $10^6$ | $(10^{17})$ | 1 | 1 |
| AGN (3C279) | $10^{48}$ | $10^{15}$ | $10^{16}$ | 100 | 3 |
| GRB | $10^{51}$ | $10^6$ (?) | $10^8$ | $10^{13}$ | 100 |

[a]pulsed component

that the emission is not isotropic. In the case of a pulsar we know this independently by the very fact that we observe *pulsed* emission from a rotating beam. If the solid angle subtended by the beam is small, i.e. $\Omega \ll 4\pi$, the power emitted is only $(\Omega/4\pi)10^{36}$ erg s$^{-1}$. Additionally, the photon-photon scattering angles are constrained to a narrow band increasing the threshold energy Eq.(3) and decreasing the pair production cross section Eq.(4). Thus, the paradox is removed. The physics responsible for the intrinsic anisotropy is intimately linked with the Goldreich-Julian limited vacuum inside the pulsar magnetosphere where large-scale electric fields can be maintained accelerating particles in particular directions. If ambient plasma would shortcut the potential drops between open field lines, the emitting particle distributions would immediately isotropize. Thus, this elegant way out of the paradox is blocked for AGNs which we believe are powered by accretion of high-conductivity plasma onto black holes. It is also blocked for GRBs where the energy release greatly exceeds the magnetic field energy.

AGNs: The rapid variability on time scales $\Delta t < 1$ day characterizing gamma ray emitting AGNs seems to indicate emission from a scale of the order of $10 - 100r_\mathrm{S}$. A strong rotating magnetosphere cannot be ascribed to the black hole itself by virtue of the 'no-hair' theorem, but very likely it is carried by plasma *accreting* onto the black hole. Matter can only spiral inward towards the black hole, if its angular momentum is extracted. Angular momentum transport could occur through a rotating magnetized wind ejected along the rotation axis (Blandford & Payne 1982, Pudritz 1985, Ferreira & Pelletier 1993) which is observed in about 10% of all AGN as a powerful radio jet. To date, there is no evidence for the competing mechanism of radial angular momentum transport in an accretion disk (Kinney 1994). The Alfvén point on each streamline (the corotation radius) would have to be at $\sim 10 - 100r_\mathrm{S}$ for an ejection/accretion rate ratio of $\dot M_\mathrm{e}/\dot M_\mathrm{a} \sim 1/100$. Since the released energy is imparted on the small amount of mass ejected through the wind, the energy per particle can asymptotically attain very high values, i.e.

$$\Gamma = E/M_\mathrm{e}c^2 \simeq L/\dot M_\mathrm{e}c^2 \simeq 100\eta_\mathrm{a} \simeq 10 \qquad (7)$$

where $L = \eta_\mathrm{a}\dot M_\mathrm{a}c^2$ denotes the accretion luminosity with efficiency $\eta_\mathrm{acc} \simeq 0.1$ in the Schwarzschild metric of a non-rotating black hole. In a Kerr metric, the efficiency $\eta_a$ can be significantly higher. *Relativistic bulk motion is required, if a tenuous magnetized wind carries away the angular momentum of the accretion flow.* Due to the pinching Lorentz force by the self-produced toroidal magnetic field, the wind can collimate to a jet flow emerging from the central engine as a very narrow cone. However, the jets are believed to be very different from pulsar beams: The particles and photons in the comoving frame of the jet can be assumed to be isotropic due to rapid pitch-angle scattering off magnetic field fluctuations. Beaming of the radiation pattern arises solely by bulk relativistic motion with Lorentz factor $\Gamma \gg 1$ in a direction at an angle $\theta$ with the line of sight. From phase space covariance, we obtain the observed flux density (Rybicki & Lightman 1979)

$$S_{\rm obs}(\nu) = \delta^{3+\alpha} S(\nu) \tag{8}$$

where $S(\nu)$ denotes the flux density in the comoving frame and where $\delta$ denotes the Doppler factor given by

$$\delta = \frac{1}{\Gamma(1 - \beta \cos\theta)} \tag{9}$$

For the special angle $\cos\theta = \beta$ we obtain $\delta = \Gamma$. The observed variability time scale is given by

$$\Delta t_{\rm obs} = \Delta t / \delta \tag{10}$$

leading to underestimates of the source size $r \leq c\Delta t$. Inserting the photon density of a source at distance $D$

$$n_\gamma = \frac{2\pi}{\hbar c} \left(\frac{D}{r}\right)^2 \int S(\nu) d\ln\nu = \delta^{-(5+\alpha)} n_{\rm obs} \tag{11}$$

into Eq.(5) yields the optical depth at the source

$$\tau_{\gamma\gamma,{\rm o}} = \delta^{-(4+\alpha)} \tau_{\gamma\gamma,{\rm o}}({\rm obs}) \tag{12}$$

Typically, $\alpha$ has values from 1/2 to 1 and the optical depth is reduced by $\sim \Gamma^{-5}$. Thus, for $\Gamma > \Gamma_{\rm o} \sim 3$ the photon-photon pair production optical depth decreases below unity. As a matter of fact, the maximum energy of blazars detected by EGRET is at least $\sim 10$ GeV requiring $\tau_{\gamma\gamma,{\rm o}} \leq 10^{-4\alpha}$ or $\Gamma \sim 8 - 16$ in sources like 3C279 which is in accord with the simple estimate Eq.(7) for a jet with $\dot{M}_{\rm e}/\dot{M}_{\rm a} \sim 1/100$.

GRBs: Again referring to Tab.1, the gamma ray absorption problem seems to be the worst in the case of gamma rays bursts. Although they emit only a short pulse of gamma rays with a typical duration of $\sim 1$ s, the observed fluence $F \sim 10^{-7}$ erg cm$^{-2}$ (Hartmann 1995) implies a luminosity of

$$L \sim 10^{51} (D/100 \text{ Mpc})^2 (F/10^{-7} \text{ erg cm}^{-2})/(t_{\rm grb}/1 \text{ s}) \text{ erg s}^{-1} \tag{13}$$

Even a halo origin at $D \sim 100$ kpc reducing the GRB luminosity by six orders of magnitude compared to cosmological GRBs does not resolve the paradox. For a cosmological distance $D \sim 100$ Mpc, which is more in line with the observed isotropy and apparent brightness distribution of GRBs (Tegmark et al. 1995), the total energy release in photons $E \sim 10^{51}$ erg is a tantalizing 1% of the gravitational energy of a collapsing solar mass. In a supernova, the energy of the collapsing core is released into a mass shell with $M \sim 10 M_\odot$ provided by the surrounding *massive* star. Thus, in the latter case the ejecta cannot reach relativistic velocities. A more or less *naked* collapse is required to obtain $\Gamma > 1$, e.g. due to the coalescence of two neutron stars (Mèszàros & Rees 1993a). Then the energy release occurs in the low density region perpendicular to the orbital plane of the coalescing or accreting neutron star binary where only a small baryon pollution is expected. Primarily the energy goes into $\nu + \bar{\nu}$ pairs and it has been estimated that 1% of the neutrinos annihilate yielding electron/positron pairs, the relativistic 'fireball'[4]. The Thomson thick fireball is highly super-Eddington, i.e. $L/L_{\rm E} \simeq 10^{13}$, and it is therefore subject to rapid expansion. During adiabatic expansion, the available energy is very soon consumed by the few polluting baryons which store the energy as bulk kinetic energy (Mèszàros et al. 1993). Radiative losses during these stages of the explosion are very small. It is the formation of a collisionless shock when the mass of the swept-up interstellar matter equals the exploding mass that non-thermal radiative losses become important. This is analogous to supernova remnants, with the difference that the fireball explosion is relativistic. The optical depth for a relativistic, spherically expanding shell is given by

$$\tau_{\gamma\gamma,{\rm o}} = \Gamma^{-(4+\alpha)} \tau_{\gamma\gamma,{\rm o}}({\rm obs}) \tag{14}$$

and the observed variability time scale by

$$\Delta t_{\rm obs} = \Delta t / \Gamma^2 \tag{15}$$

Hence the criterion $\tau_{\gamma\gamma,{\rm o}} \leq 1$ requires a bulk Lorentz factor

$$\Gamma_{\rm o} = \frac{E}{M_{\rm e} c^2} = 10^{\frac{13}{4+\alpha}} \sim 10^2 - 10^3 \tag{16}$$

where $M_{\rm e}$ denotes the conserved baryonic rest mass. The size of the expanded fireball is $r_{\rm d} = \Gamma^2 c \Delta t \sim 10^{16}$ cm which is much larger than the initial compact object. Its small optical thickness is in accord with the observed non-thermal spectra. By contrast, the precursor event is extremely short, optically thick and less bright than the optically thin relativistic blast wave (Mèszàros & Rees 1993b, Mèszàros et al. 1994). A GRB should therefore be viewed as a relativistic supernova remnant.

---

[4]For a criticism, see Janka & Ruffert, MPA-Report 909

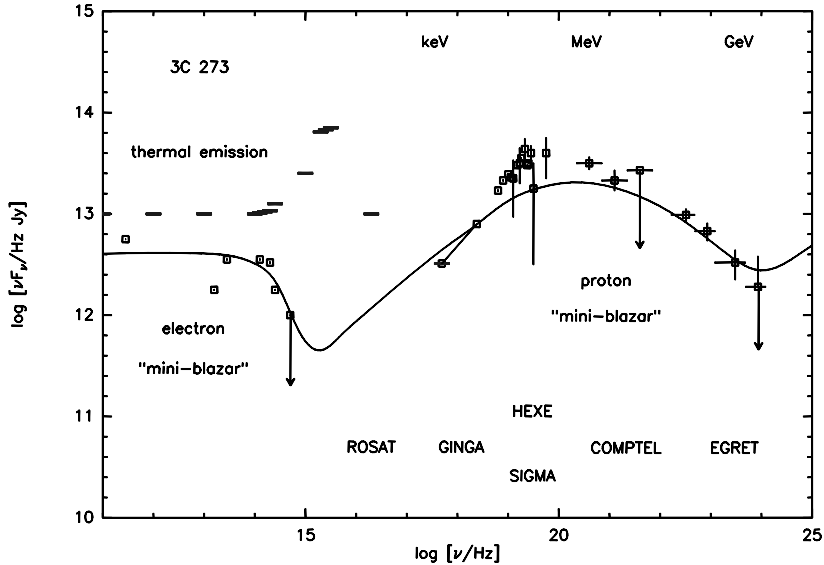

Figure 1: 3C273: Proton-initiated cascade spectrum fitted to the non-thermal jet spectrum (Mannheim 1993b). In addition to the jet spectrum, the data (Lichti et al. 1995) show thermal emission in the FIR, UV and soft X-ray bands presumably emitted by an accretion flow and a surrounding dust torus. Above ∼100 GeV, the gamma ray spectrum cuts off sharply due to pair absorption in the infrared radiation field of the dust torus

## 3 Relativistic jets and accreting black holes in AGNs

The most detailed, but still controversial, understanding of the radiation mechanisms responsible for gamma ray emission from compact objects has been achieved for pulsars (e.g., Cheng et al. 1986, Daugherty & Harding 1982, Gallant & Arons 1994) and AGNs (e.g., Blandford & Levinson 1995, Dermer & Schlickeiser 1993, Henri et al. 1993, Sikora et al. 1994). In this Lecture, emphasis is on the author's hadronic emission model for AGNs. The non-stellar activity characteristic for AGNs basically comes in two manifestations, (i) a UV-to-soft-X-ray bump indicating thermal emission and (ii) a broad non-thermal continuum extending from radio frequencies to gamma rays. The relative strength of the two emission components varies by orders of magnitude between members of different classes of AGNs. Radio-quiet AGNs show practically only the thermal emission component (plus a thermal FIR bump due to reprocessing by heated dust and a weaker X-ray component which could be either thermal or non-thermal), whereas radio-loud AGNs of the blazar subtype only show the non-thermal component. Fig.1 shows the mixed thermal/non-thermal spectrum of the quasar 3C273.

RADIO-LOUD AGNs: Morphologically, at a sufficient angular resolution powerful radio emission is related to radio jets. For very large Alfvénic Mach numbers $M_A \gg 1$, the rotating hydromagnetic wind from the accretion flow onto a supermassive black hole can emerge from the central parsec of the AGN propagating out of the host galaxy and deep into intergalactic space (Camenzind 1990). Estimating the kinetic power of radio jets by applying strong shock jump conditions to the terminal shocks (hot spots) of jets where they hit the intergalactic medium (Roland et al. 1988, Meisenheimer et al. 1989), by computing the radio age (Rawlings & Saunders 1991) or by synchrotron-self-Compton limits (Celotti & Fabian 1993, Daly et al. 1996), generally yields values of the order of the Eddington luminosity for black holes in the mass range $10^8 - 10^{10} M_\odot$. Therefore, it is evident that the jets *must* emerge from the central engine, no other comparable power supply is available. The emergence of the jet from the central region has the fascinating implication that the plasma from the vicinity of the black hole can be probed on scales which are observationally resolvable using very long baseline interferometry (VLBI). One of the major results is that knots of enhanced surface brightness in the radio jets can move at *superluminal* speeds (e.g., Porcas 1987, Begelman et al. 1994). The most natural and robust explanation of this observation is that of relativistic motion of the emitting plasma at a small angle to the line of sight which can give apparent speeds $v_{j,app} \sim \Gamma v_j$ when projected onto the plane of the sky. Another consequence of relativistic bulk motion is the beaming of the photon flux into a cone of opening angle $1/\Gamma$, see Eq.(8). Consequently, the appearance of a radio source must depend strongly on the angle between the radio axis and the line of sight. As a matter of fact, the zoo of radio-loud AGNs can be unified to the species of powerful (FRII) radio galaxies (seen edge-on) appearing as flat-spectrum radio quasars (Fig.1) when seen pole-on and to the weaker FRI radio galaxies appearing as BL Lacertae objects (Fig.2) for a pole-on orientation (Urry & Padovani 1995). In addition to the orientation-dependence of the radio emission, the nuclear emission line regions and the UV bump also vary with angle (Antonucci 1993). An obscuring torus hides the central parsec within $\sim 45°$ from the edge-on orientation. In the spectral energy distribution, the importance of the non-thermal continuum increases with increasing pole-on orientation. All radio-loud AGNs with their jet axes oriented at small angles to the line of sight are summarized under the name 'blazars' due to their variable, polarized synchrotron spectra from a boosted relativistic jet. The major unanswered questions affecting the nature of the gamma ray emission mechanism are:

(1) What is the composition of the jet plasma (Poynting flux, $e^\pm$ pairs, solar abundance plasma, enriched plasma, something else)?

(2) At which distance to the black hole is the emission produced?

(3) What is the comoving-frame ratio of magnetic field energy density and radiation energy density?

(4) What is the acceleration mechanism producing the ultrarelativistic particles in the comoving frame responsible for the non-thermal radiation?

In the baryon-induced cascade model for the gamma rays presented in this Lecture, it will be assumed that (1) the plasma contains protons and electrons, (2) the distance of the non-thermal source in the jet is large enough (∼ 1 lyr) to make the locally produced synchrotron photons the dominant target contributing to the gamma ray opacity, (3) the magnetic field energy density is assumed to dominate over the radiation energy density and, finally, (4) shock acceleration is the acceleration mechanism at high energies. These assumptions lead to a consistent physical picture and a predictive model (Sect.4). Let us discuss some arguments in favor of the above assumptions.

**Ad (1)** If the jet originates as a hydromagnetic wind ejected by the accretion flow, the plasma flowing along the streamlines is of the same composition as the plasma accreted onto the black hole. Protons should be the dominant baryonic species in this plasma. Alternatively, if the jet originates as Poynting flux produced by a maximally rotating black hole, the Poynting flux would be converted to a relativistic pair plasma close to the ergosphere by photon-photon interactions in the UV/X-ray radiation field of the optically thick accretion flow. As in the case of fireballs, however, the unavoidable polluting baryons ultimately receive the released energy during an adiabatic expansion, while the pairs would cool and annihilate. Nevertheless, the absence of Faraday depolarization seems to argue in favor of a pure pair plasma in which Faraday depolarization is absent (Jones 1988). However, in an ordinary e-p jet plasma which is relativistically hot, with temperatures ∼ 100 MeV, Faraday depolarization would also be absent. Such temperatures are expected behind shock waves propagating down the tenuous jet flow (Blandford & McKee 1976). Runaway pair production by collisional processes does not occur, since the density in the jet plasma is too low (particle-wave interactions dominate).

**Ad (2)** Dissipation is certainly important at the base of the jet where an entropy generating demon loads particles onto the field lines. The heated jet base would act as a comptonizing screen through which the UV photons from the optically thick accretion flow have to pass. Unsaturated comptonization of the UV photons by the hot jet base produces power law soft X-ray excesses which are commonly observed in AGNs (Walter & Fink 1993). For the particle energy distributions to develop power law tails reaching ultrarelativistic energies, it is necessary that the density in the jet becomes very low and that *collisionless* shocks form. While the formation of shocks due to magnetic flare activity at the base of the jet seems to be unescapable, the rather high density $n \sim 10^{10} - 10^{12}$ cm$^{-3}$ and consequently large Coulomb collisional rate tends to thermalize the injected energy ($t_{\rm coul} < r_{\rm j}/c$ for $n > 10^{10}$ cm$^{-3}$ at $T = 10^9$ K

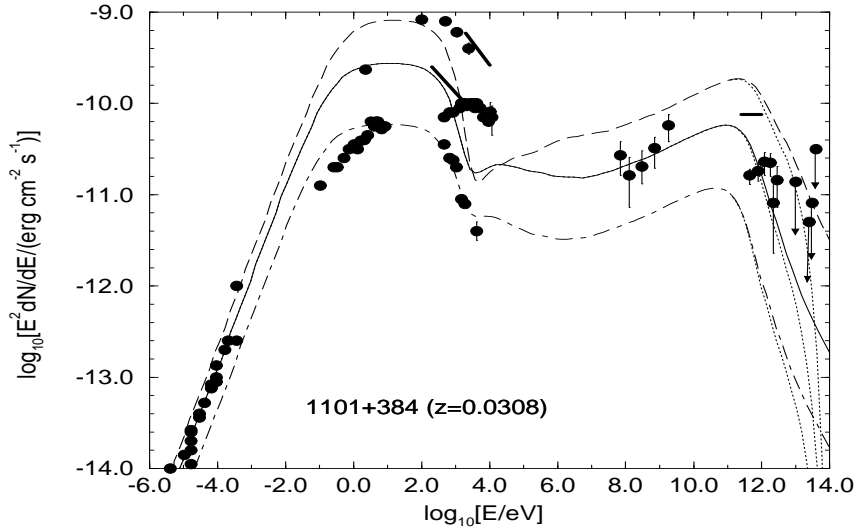

Figure 2: Mrk421: Proton blazar fits to the highly variable broadband continuum obtained by changing the angle between the velocity of the emitting plasma in the jet and the line of sight by a few degrees (shocks traveling in different directions?). Note that Mrk421 is detected over 18 orders of magnitude in frequency

and $r_\mathrm{j} = 100 r_\mathrm{S}$, Mannheim & Schlickeiser 1994). In fact, flare-induced shock dissipation could be the major heating mechanism of the jet base presumably responsible for soft X-ray emission. Therefore, the most likely site of the gamma ray emission region in the jet is far away from the base of the jet when $n \ll 10^{10}$ cm$^{-3}$. Shocks would be generated behind the magnetic nozzle provided by the self-confining toroidal field (van Putten 1995) when the jet expands into the steep pressure gradient of the interstellar medium of the host galaxy. The distance of this region to the central black hole could be very large, i.e. $r \gg 1$ la.

**Ad (3)** The most direct measurement of the ratio $a = u_\gamma/u_B$ comes from the hot spots of Cyg A (Harris et al. 1994) yielding $a \leq 10^{-2}$, cf. Fig.3. The idea behind the method is that X-ray observations limit the comoving-frame ratio of the photon-to-magnetic energy density, since the observed synchrotron emission from the jets implies synchrotron-self-Compton X-rays due to the Compton scattering of synchrotron photons with energy $\epsilon_\mathrm{syn}$ off the same relativistic electrons with Lorentz factor $\gamma_\mathrm{e}$ which produced them in the first place. In the Thomson scattering regime ($\epsilon_\mathrm{syn}\gamma_\mathrm{e} < m_\mathrm{e}c^2$), the characteristic photon energy of inverse-Compton scattered photons is given by

$$\epsilon_\mathrm{ic} \sim \epsilon_\mathrm{syn}\gamma_\mathrm{e}^2 \qquad (17)$$

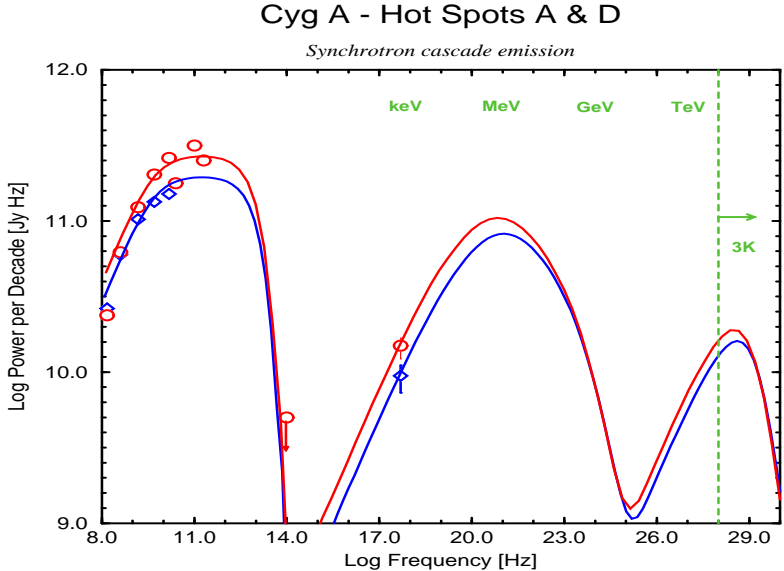

Figure 3: Cyg A: Proton-initiated cascade fits to the hot spot spectra. The radio-to-optical spectrum is synchrotron emission by accelerated electrons. The X-ray-to-gamma-ray spectrum is unsaturated synchrotron cascade emission by accelerated protons assuming $u_\mathrm{p} = 40 u_\mathrm{e}$ and $B = 3 \times 10^{-4}$ G

which easily reaches the X-ray range. The SSC X-ray luminosity can be approximated by

$$L_\mathrm{ssc} \sim \frac{u_\gamma}{u_B} L_\mathrm{syn} \qquad (18)$$

In a stationary state, the photon energy density should not exceed the magnetic field energy density, since the particles are required to scatter off the magnetic field fluctuations $u_{\delta B} \sim u_B$ to be accelerated by a statistical Fermi (e.g., Drury 1983) or drift mechanism (Webb et al. 1983, Begelman & Kirk 1990). Thus, if $u_\gamma > u_B$ the emitting particles would escape from the magnetic confinement. Moreover, the back-reaction of the inverse-Compton cooling on the electron energy distribution would act to equilibrate the energy losses over a broad energy band leading to continua smoothly joining the optical and X-ray ranges by an $\alpha = 1$ power law which is apparently not the case for the Cyg A and other hot spots. Assuming now $u_\gamma \propto r^{-2}$ and a tangled magnetic field $B \propto r^{-1}$ (adiabatic expansion in a conical jet implies $B_\perp \propto r^{-1}$ and $B_\parallel \propto r^{-2}$) leads to $a = constans$ implying that synchrotron emission is the dominant cooling mechanism for the relativistic electrons in the jet. This is in accord with the fact that the beamed hard X-ray luminosity from nuclear jets typically does not exceed the synchrotron luminosity Eq.(18). Nevertheless, it could be possible that the external radiation field energy density exceeds the magnetic field energy density close enough to the thermal source of UV photons. This would not be the case in angular mo-mentum carrying jets where an initial equipartition $u_\gamma = u_B$ would lead to an ever-increasing domination of the magnetic field in the jet with increasing distance[5]. Thus models in which the dominant energy loss is not due to synchrotron emission imply the assumption that the jet magnetic field is initially below equipartition and that it is increased by dynamo action in the hot spots. The importance of the rotating jet as an angular momentum transport structure is then diminished accordingly.

**Ad (4)** Shock acceleration theory has been applied successfully to hot spot spectra at the kpc scale and the local spectra produced by knots in the jet at smaller scales (VLA, VLBA, VLBI) seem to be quite similar. Moreover, there is a detailed correspondence of shock-in-jet models with the observed variability (Hughes et al. 1989, Quian 1991) and from numerical simulations of supersonic jets crossing steep pressure gradients shock formation seems unavoidable. Periodic shock structures can be formed by magnetic nozzles (van Putten 1996). Moreover, shock acceleration is known to work from *in situ* experiments in the solar system and it is the leading contender explaining the acceleration of cosmic rays at supernova remnants. The major drawback is the lack of a detailed understanding of how *relativistic and oblique* shocks accelerate particles. A large computational and theoretical effort is necessary to improve the situation.

RADIO-QUIET AGNs: An explanation of the absence of strong gamma ray emission from radio-quiet AGNs (Maisack et al. 1995) is a substantial requirement for any model explaining gamma rays from radio-loud AGNs. In radio-quiet quasars and Seyfert galaxies, the thermal big blue bump dominates the bolometric luminosity by far (Sanders et al. 1989). Powerful radio jets are generally not present in these objects, but faint bipolar radio morphologies seem to be quite frequent. As in the case of radio-louds, accretion onto a black hole and angular momentum transport by a rotating MHD wind constitutes a viable explanation. The competing idea of a thin accretion disk where angular momentum is transported radially outwards by viscous friction between turbulent eddies is in conflict with observations (Kinney 1994). E.g., the optical emission should come from much larger radii than the UV and soft X-ray emission. Nevertheless, flux variations occur quasi-simultaneously from the optical to X-ray range (Clavel et al. 1992). Coupled accretion/ejection structures could ameliorate the problem in many ways (Mannheim 1995b). On the other hand, the absence of radio (and gamma ray) emission from radio-quiets is often invoked as counter-evidence against a nuclear wind. However, if the Alfvénic Mach number of the rotating magnetized wind is $\sim 1$, Kelvin-Helmholtz instabilities grow rapidly over a length scale of order $\sim 10 r_{\rm S}$. As a consequence, the vertical wind with large opening angle $\sim 1/M_{\rm A}$ becomes turbulent (Mannheim 1995b). In this way, the formation of strong shocks propagating down the jet producing powerful radio emission is avoided explaining 'radio-quietness'. The wind in its outer parts is colder and only partially ionized providing a warm absorber seen in the X-ray spectra of many AGNs. As the formation of the hydromagnetic wind is certainly accompanied by dissipative processes, the base of the tenuous wind close to the putative black hole must be heated to very high temperatures, e.g. by magnetic reconnection. Any line of sight to the UV-emitting optically thick plasma of the accretion flow must pass through this hot wind base, thus producing soft X-ray power law spectra by unsaturated Comptonization of the UV photons – there is no fundamental difference between radio-quiets and radio-louds in this respect (the soft X-ray spectra in radio-louds should be slightly flatter). Rapid correlated variations on a dynamical time scale occur naturally in this scenario (e.g., due to coronal mass ejections). Following the arguments in Sect.2, radio-quiet AGNs should represent disk/jet systems where the ejected mass loss rate larger than in sources with relativistic jets and consequently their jets are non-relativistic. Any non-thermal emission associated with particle acceleration in the wind (e.g., stochastic acceleration Henri & Pelletier 1991, Dermer et al. 1995, Mannheim 1995b) *occurs well within the gamma ray photosphere of the thermal UV/X-ray radiation field ($\tau_{\gamma\gamma} > 1$)*, since

$$r_{\rm o}(\tau_{\gamma\gamma} = 1) = 10 r_{\rm S} \epsilon_{-1} \tau_{-2} \left(\frac{E}{10 m_{\rm e} c^2}\right) \tag{19}$$

assuming $\alpha_{\rm x} = 1$, $L_{\rm x} = 0.1\epsilon_{-1} L_{\rm E}$ and the scattered isotropic radiation field $L_{\rm x,iso} = 0.01 \tau_{-2} L_{\rm x}$. Whereas hard X-rays are reflected by Thomson scattering off the cold accretion flow, the gamma rays are absorbed heating the flow. Therefore, radio quiet AGNs cannot be powerful gamma ray emitters[6]. The absence of a diffuse gamma ray bump with a power comparable to the 30 keV X-ray bump in the diffuse background radiation field supports these conclusions.

---

[5] For comparison, a thermal radiation field of the same power as the jet in a typical FR II radio galaxy with $L \sim 10^{46}$ erg s$^{-1}$ would have an energy density of $L/(4\pi D^2 c) \sim 3 \times 10^{-13}$ erg cm$^{-3}$ at a distance of $D \sim 100$ kpc, whereas the magnetic field $B \sim 10^{-4}$ G has $B^2/(8\pi) \sim 4 \times 10^{-10}$ erg cm$^{-3}$ in hot spots.

## 4  Baryon-initiated cascades

The spectra of radio jets are characterized by a polarized radio-to-X-ray featureless continuum which is commonly interpreted as synchrotron emission by relativistic electrons with energies of maximally $\sim 100$ GeV. Typically, the spectral slopes of *local* emission regions in the jet (knots, hot spots) are $\alpha \sim 0.5$ ($S_\nu \propto \nu^{-\alpha}$) in the radio-to-submm range and $\alpha \sim 1$ in the infrared. Further steepening is observed in the the optical-to-soft-X-ray range. At hard X-rays, the spectra seem to rise again possibly indicating synchrotron-self-Compton emission. The origin of hard X-rays and gamma rays cannot be resolved spatially, however, their association with jets seems to be beyond doubt. In greatest detail, the radio-to-X-ray spectra have been studied for hot spots in the jets of FR II radio galaxies at a scale of ∼kpc. By two independent methods the magnetic field strengths in hot spots could be inferred yielding $B_{hs} \sim 3 \times 10^{-4}$ G (Meisenheimer et al. 1989). Closer towards the core, the spectra of knots in the jet show a low-frequency synchrotron-self-absorption turnover increasing in frequency with decreasing distance from the core. *Apparently, regions of enhanced surface brightness density along the jet represent scaled-down versions of hot spot spectra in regions of larger magnetic field ($B_j \sim 0.1 - 1$ G).* Knots at very small scales ∼ 0.1 − 1 pc superimpose to a more or less conical emission region. As a result of the superposition of increasingly self-absorbed spectra, the nuclear jet obtains a flat radio spectrum (Kellermann & Pauliny-Toth 1981). VLBI observations of superluminal motion in nuclear jets indicate relativistic bulk motion, whereas the more distant parts of the jet (hot spots) are very likely at least moderately relativistic resulting from a general slowing down of the jet. Hence it follows that in the relativistic part of the jet the hot spot spectra are boosted by the Doppler factor $\delta \sim 3 - 10$ in frequency and by $\delta^4$ in luminosity. Therefore, the nuclear jets appear as extremely bright and variable blazars when viewed at a small angle of the jet axis to the line of sight. On the other hand, when the observer is outside of the boosting cone with opening angle $1/\Gamma$, the nuclear jet is greatly diminished in brightness and the otherwise washed out more distant and isotropically emitting parts of the jet (hot spots, lobes) become visible. In spite of their brightness, blazars are very difficult to diagnose because of the relativistic projection effects, the lack of spatial resolution at a high dynamical range and the complex environment in the center of the active galaxy. Therefore, the best laboratory to study the non-thermal radiation processes in jets are the isolated hot spots of FR II radio galaxies, such as Cyg A.

---

[6] There is a possibility that soft gamma rays of energy $\epsilon \sim$ few MeV are emitted with an anisotropic distribution preferentially towards the disk. In this case, we expect weak gamma ray emission from Seyfert 2 galaxies which we believe are AGNs seen near the plane of the accretion flow.

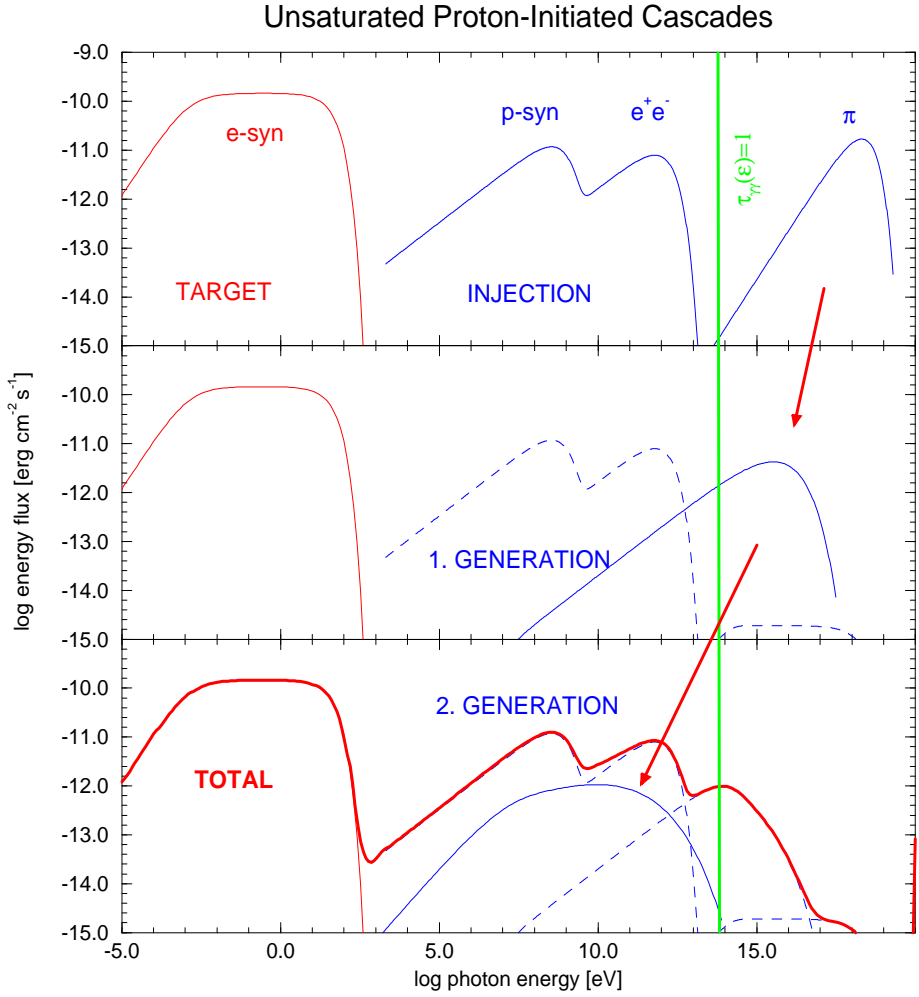

Figure 4: Sketch of the PIC mechanism: Shock-accelerated electrons produce radio-to-X-ray synchrotron radiation (target). Shock-accelerated protons cool in the synchrotron target photons and magnetic field producing pions, pairs and synchrotron emission (for the sake of demonstration, all three secondary luminosities have been made equal). The neutral pions decay producing ultrahigh energy gamma rays (*top panel*). Electromagnetic power injected into the acceleration region above the energy $E^*$ where $\tau_{\gamma\gamma} = 1$ is absorbed creating relativistic pairs. These pairs initiate the first generation of cascade gamma rays via synchrotron emission (*middle panel*) and the cycle repeats. In the above example, already the second generation of gamma rays emerges almost completely below $E^*$ terminating the cascade (*bottom panel*). The resulting total spectrum is given by the superposition of the individual absorbed cascade generations and the primary electron synchrotron spectrum.

What is the origin of the relativistic synchrotron-emitting electrons? For radio jets which reach out to distances of ∼Mpc, it is impossible that the relativistic electrons (or positrons) originate in the active nucleus. Before they arrive at the hot spots, they would loose their energy by inverse-Compton scattering off the 3K-microwave background. Therefore, *in situ* acceleration in the hot spots is required. The observed power law energy distributions $dn_e \propto \gamma_e^{-2} d\gamma_e$ argue for a statistical process. Diffusive (e.g., Drury 1983) or drift transport (Begelman & Kirk 1990) of relativistic particles across a strong collisionless shock are the leading contenders among the possible *in situ* acceleration mechanisms. A major drawback of the theory is the lack of an efficient pre-acceleration mechanism. Resonant scattering off Alfvén waves requires that the particle momenta exceed a threshold momentum which is already relativistic in the case of electrons. Pre-acceleration of electrons out

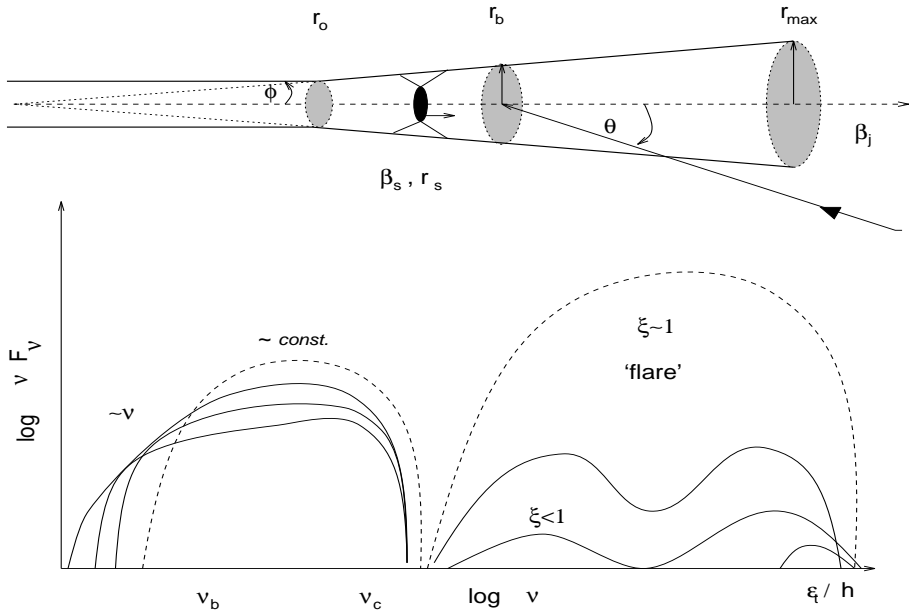

Figure 5: Sketch of the blazar spectrum due to emission from varius nuclear jet radii. Injection of a shock at small radii propagating along the jet increases the acceleration efficiency and maximum energies producing a 'flare'. The parameter $\xi$ denotes the ratio of the proton maximum energy to the limit in Eq.(26). For fresh shocks, $\xi \sim 1$ while nonlinear shock evolution decreases the acceleration efficiency yielding $\xi < 1$

of the thermal pool could be achieved by resonant interaction with Langmuir and Whistler waves, or by turbulent reconnection of magnetic field lines. For protons, the injection problem is less problematic, since a rather large number of protons from the Maxwell tail of the thermal distribution is available above the Alfvénic threshold (Malkov & Völk 1995). Moreover, the Larmor radius of a gyrating proton is much larger than that of an electron increasing the acceleration efficiency, since the shock front is likely to be broadened to a nonlinear structure. The Larmor radius sets a typical scale for pitch-angle scattering, so traversing the shock on a single scattering (which is essential to gain energy from the Lorentz transformation between the upstream and downstream side of the shock) should be easier for protons than for electrons. In the local cosmic ray gas, the energy density of relativistic protons above GeV exceeds that of the relativistic electrons by a factor of $\sim 100$, although this does not prove that proton acceleration is generally more efficient at high energies. A first-principle prediction of the p/e energy density ratio $\eta = u_\mathrm{p}/u_\mathrm{e}$ is not possible at the current status of the theory.

Historically, protons have generally been neglected as a radiative agent. Their main energy loss at high energies is pion production by collisions with matter as indicated by the $\pi^\circ$ bump at 70 MeV in the gamma ray emission from the galactic disk due to the interaction of cosmic rays with the interstel-lar matter. However, in radio jets the matter density is extremely low, viz. $\sim 10^{-4}$ cm$^{-3}$ at $r \sim 1$ kpc. The p-p pion luminosity for a $L_{\rm syn} \sim 10^{45}$ erg s$^{-1}$ jet amounts only to the negligible value

$$L_{\rm pp} \sim \frac{u_{\rm p} V}{t_{\rm pp}} \sim \eta L_{\rm syn} \frac{t_{\rm syn}}{t_{\rm pp}} \sim 10^{-8} \eta L_{\rm syn} \qquad (20)$$

The negligible energy losses of relativistic protons imply that they must reach very high energies in a statistical process balancing escape and acceleration. In fact, the energies reach up to values where the acceleration time scale equals the energy loss time scale (unless the Larmor radius would exceed the jet radius). As pointed out by Biermann & Strittmatter (1987) and Sikora et al. (1987), protons should therefore reach energies where the acceleration is halted by photo-production of pions and pairs. The processes

$$p + \gamma \rightarrow \pi^\circ + p \ , \quad \pi^+ + n \qquad (21)$$

$$p + \gamma \rightarrow e^+ + e^- + p \qquad (22)$$

have effective inelastic cross sections $\sigma_\pi \sim m_\pi/m_{\rm p} \alpha_{\rm f} \sigma_{\rm pp} \sim 10^{-28}$ cm$^2$ and $\sigma_{e^\pm} \sim 2m_{\rm e}/m_{\rm p} \alpha_{\rm f} \sigma_{\rm T} \sim 10^{-30}$ cm$^2$ (Blumenthal 1970). The dimensionless threshold energies for a head-on collision with a low energy synchrotron photon of energy $\epsilon$ are

$$\gamma_{\rm p, th, \pi} = \frac{m_\pi c^2 (1 + \frac{m_\pi}{2m_{\rm p}})}{2\epsilon} \qquad (23)$$

and

$$\gamma_{\rm p, th, e^\pm} = \frac{2 m_{\rm e} c^2}{2\varepsilon} \qquad (24)$$

Typical values for optical target photons $\epsilon \sim 1$ eV are $\gamma_{\rm p, th, \pi} \sim 10^8$ and $\gamma_{\rm p, th, e^\pm} \sim 10^6$. In a power law target radiation field with $\alpha = 1$ (IR-to-optical), pair and pion production contribute equally to the proton cooling. The proton maximum energy is given by equal acceleration and energy loss time scales. Since the shortest acceleration time scale is obtained for Bohm diffusion ($t_{\rm acc} \geq r_{\rm L}/c$), we arrive at the estimate

$$\gamma_{\rm max} \leq 2 \times 10^{11} \left( \frac{B}{3 \times 10^{-4} \text{ G}} \right)^{-\frac{1}{2}} a^{-\frac{1}{2}} \qquad (25)$$

Additionally, we require that the Larmor radius does not exceed the jet radius (Bell 1978) obtaining

$$\gamma_{\rm max} < 3 \times 10^{11} \left( \frac{B}{3 \times 10^{-4} \text{ G}} \right) \left( \frac{r_{\rm j}}{1 \text{ kpc}} \right) \qquad (26)$$

Therefore, proton acceleration in hot spots ($a \sim 0.01$) is bounded to values of order $\gamma_{\rm max} \sim 3 \times 10^{11}$ by geometrical constraints. In the comoving frame of nuclear jets ($B \sim 1$ G, $r \sim 0.1$ pc) we obtain $\gamma_{\max} \sim 3 \times 10^{10}(a/0.01)^{-1/2}$ for Bohm diffusion and the Larmor radii of the most energetic particles remain an order of magnitude below the jet radius.

The proton secondary pairs and pions are the origin of non-thermal high energy radiation from jets. The pairs cool instantaneously via synchrotron radiation (inverse-Compton radiation is down by a factor $a \sim 0.1 - 0.01$), the pions decay into gamma rays via $\pi^\circ \to \gamma + \gamma$ and produce further positrons via $\pi^+ \to e^+ + 3\nu$. We have seen in Sect.2 that the optical depth for gamma rays of energy $E$ is proportional to $E^\alpha$. Since $E$ reaches very large values for gamma rays of secondary origin, the optical depth inevitably exceeds unity. The gamma rays are absorbed within the radio jet producing pairs, which themselves produce further gamma rays etc. The cycles terminate when the gamma ray energy of the n-th generation decreases below the energy $E^*$ where $\tau_{\gamma\gamma}(E^*) = 1$. The process is sketched in Fig.4 (Mannheim et al. 1991). Note that the thermal pair yield, and hence the flux of the annihilation line, is very low, since unsaturated cascades produce only highly relativistic pairs (Svensson 1987).

In the most simple model of the cascade, local power law injection of secondary pairs and gamma rays as well as primary electrons into a spherical region is assumed. The emerging photon spectrum corresponds to a stationary solution of the coupled set of kinetic equations for the pair energy distribution $n(\gamma, t)$ and the photon energy distribution $m(x, t)$ at dimensionless energy $x = \epsilon/m_e c^2$ given by

$$\frac{\partial n(\gamma, t)}{\partial t} + \frac{\partial}{\partial \gamma}[\dot{\gamma}_{\text{syn}} n(x, t)] + \int_\gamma^\infty n(\gamma', t) \int_1^\infty m(x, t) G_{\text{com}}[\gamma, \gamma', x] dx d\gamma'$$
$$- n(\gamma, t) \cdot \int_1^\infty \int_0^\infty m(x, t) G_{\text{com}}[\gamma, x, \gamma'] dx d\gamma' + \frac{n(\gamma, t)}{T_{\text{e,esc}}}$$
$$= q_e[m](\gamma, t) + q_e(\gamma, t) \tag{27}$$

$$\frac{\partial m(x, t)}{\partial t} + \int_1^\infty n(\gamma, t) F_{\text{syn}}[\gamma, x] d\gamma + \int_x^\infty m(x', t) \int_1^\infty n(\gamma, t) F_{\text{com}}[\gamma, x', x] d\gamma dx'$$
$$- m(x, t) \int_0^\infty \int_1^\infty n(\gamma, t) F_{\text{com}}[\gamma, x, x'] d\gamma dx' + \frac{m(x, t)}{T_{\gamma, \text{esc}}}$$
$$= -q_\gamma[m](x, t) + q_\gamma(x, t) \tag{28}$$

where the $q_i$ denote sources and sinks due to pair production and injection and the other terms denote energy losses (emissivities, resp.) due to synchrotron and Compton radiation. Diffusive escape from the acceleration region is accounted for by an escape time which is related to the lowest eigenvalue of the spatial diffusion equation. Acceleration would be encompassed by including diffusion terms and by solving the kinetic equations in the shock geometry. In Eqs.(27) and (28), the acceleration process is hidden in the power law injection terms. However, in the high-energy limit (X-rays, gamma rays) diffusion is not the dominant process for the secondary pairs.

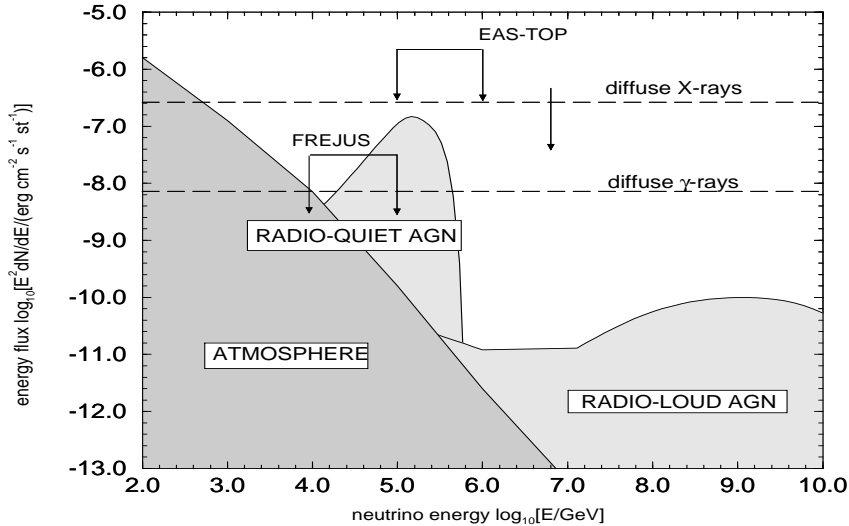

Figure 6: The cosmic diffuse neutrino background produced by AGNs. The diagonal line shows the atmospheric background due to large-angle $\mu$'s from cosmic ray interactions (Gaisser et al. 1993). Experimental limits by the Frejus and EAS-TOP collaborations are shown for comparison (see Mannheim 1995a for references). The PeV limit comes from the non-detection of contained events due to the Glashow-Weinberg resonance assuming an equal mixture of electron and muon neutrinos

In Fig.3 the model is applied to the hot spots of Cyg A where X-ray emission was detected with the High Resolution Imager of ROSAT (Harris et al. 1994). Synchrotron-self-Compton emission is neglected using a magnetic field value of $3 \times 10^{-4}$ G and $\eta = 40$. Neglecting proton-initiated-cascade (PIC) emission would yield a magnetic field of $10^{-4}$ G for $\eta = 0$. The spectral slope of the SSC emission would be close to $\alpha = 1$, whereas the PIC emission has $\alpha = 0.7$. The sensitivity of current gamma ray experiments is insufficient to test the model predictions at very high energies.

To model the entire blazar spectrum with the simple spherical proton-initiated-cascade model, we proceed as follows. The total primary electron synchrotron spectrum is modeled as a superposition of a large number of individual knots in a conical jet with the magnetic field scaling as $B \propto r^{-1}$ (Blandford & Königl 1979). For proton-initiated cascades, only target photons in the infrared-to-optical regime are important due to the threshold energy Eqs.(23),(24). The largest radiation compactness of infrared-to-optical photons comes from the jet radius $r_b$ where the spectral steepening of the local synchrotron spectrum due to cooling across the expanding emission volume occurs at the same frequency as the synchrotron-self-absorption (Blandford & Königl 1979). Cascade emission from larger radii emerges at higher en-ergies (lower value of $\tau_{\gamma\gamma}$) and has a lower flux (less target radiation), see Fig.5. At $r < r_{\rm b}$ the synchrotron efficiency of the jet is greatly reduced due to self-absorption. Therefore, modeling cascade emission by a Doppler-boosted spherical blob at $r = r_{\rm b}$ is a fair approximation. A flare due to the propagation of a strong shock along the conical jet would reach its maximum luminosity at this stage. All relevant physical parameters of the jet except for the proton/electron ratio $\eta$ can be obtained from their radio-to-X-ray spectra and morphology. Together with gamma ray measurements, the 'proton blazar' model allows for a complete diagnosis of the jet plasma.

Results from fitting the proton blazar model to a number of nearby blazar spectra are consistent with a more or less constant proton-to-electron energy density ratio $\eta \sim 100$ (Mannheim et al. 1995b). The differences in gamma ray luminosities seem to be largely due to differences in the proton maximum energies. Variability can occur on very short time scales for flattened Mach disk like geometries, inhomogeneities in the jet flow (Quian 1991), or optical depth effects of the gamma ray photosphere. Gamma rays and optical emission should vary in a correlated way. Short time lags (photo-production cooling time scale, proton acceleration time scale) are possible. The gamma ray variations have a larger amplitude than the target photon variations due to the quadratic effect that the gamma ray flux is dependent on $\gamma_{\rm p,max}$ and on $\gamma_{\rm e,max}$, whereas the radio-to-X-ray synchrotron emission is only proportional to $\gamma_{\rm e,max}$. Monitoring of 3C279 is consistent with these predictions (Maraschi et al. 1994). Due to the often sparse frequency coverage of the measurements, there is still considerable ambiguity in the diagnostic method. In addition, the blazar spectra are highly variable demanding contemporaneous data sets. It is therefore a *prime goal* to coordinate contemporaneous observational campaigns on blazars in the radio, mm, submm, FIR, NIR, optical, UV, X-ray and gamma ray ranges.

# 5 High-energy neutrinos and their detection

Decay of the charged pions produces high energy neutrinos

$$\pi^+ \to \mu^+ + \nu_\mu \to e^+ + \nu_e + \bar{\nu}_\mu + \nu_\mu \tag{29}$$

Typically, the four light particles in the final state each receive an energy of $E_{\rm p}/20$. Considering the maximum energies along the jet Eq.(25), we can expect a neutrino flux peaking in the range $10^8 - 10^{10}$ GeV. The neutrino to gamma-ray flux ratio is determined solely by decay kinematics and the relative production cross sections of charged and neutral pions (given by Clebsch-Gordan coefficients). As a result, neutrino flux predictions can be scaled to the known photon fluxes of the sources. In contrast to the photons, however, which suffer cascading, the neutrinos are always optically thin. Therefore the neutrino spectrum is the original production spectrum. Fig.6 shows the predicted neutrino spectra assuming that the diffuse X-ray background is made by radio-quiet AGNs and the diffuse gamma ray background by radio-loud AGNs. In the case of radio-quiets, it is assumed that the X-rays are non-thermal and associated to (anisotropic) pair cascades in low Mach number jets. Stochastically accelerated protons in the turbulent jets reach maximum energies of $\sim 10^7$ GeV limited by threshold pion production Eq.(23) in the UV field of the accretion flow (Colgate 1983).

Table 2: Muon events per $10^5$ m$^2$ per year and per steradian at various threshold energies $E_\nu$ for the diffuse AGN backgrounds shown in Fig.6

| $E_\nu$ [TeV] | radio-loud | radio-quiet |
|---|---|---|
| 1 | 15 | 887 |
| 10 | 3.0 | 836 |
| $10^2$ | 0.6 | 458 |
| $10^3$ | 0.1 | 0 |

The spectra thus obtained for individual AGNs are integrated over a distribution of AGNs in redshift. The limit from the Frejus proton decay experiment seems to rule out that proton-initiated cascades produce the X-rays in radio-quiets (on the assumption that the diffuse X-rays are due to AGNs). This is in accord with the fact that the general indication of non-thermal processes in radio-quiet AGNs is very poor. On the other hand, the prediction of neutrinos from radio-loud AGNs is a straightforward consequence of shock acceleration. Current flux limits are still off the predictions by orders of magnitude. However, the linearly rising neutrino-matter interaction cross section (up to the energy where the W-boson mass becomes important in the scattering propagator) and the long mean free path of the muons make it possible to build underwater or underice detectors providing sufficient sensitivity to discover the neutrinos. The H$_2$O acts as a Čerenkov medium for the relativistic muons resulting from charged current interactions between the extraterrestrial neutrinos with matter in the earth. The Čerenkov light is emitted in a narrow cone along the direction of the initial neutrino. It can be detected and the track can be reconstructed with conventional photomultiplier techniques (Gaisser et al. 1994). A class of such neutrino telescopes aiming at $10^5$ m$^2$ effective area are under construction (see references in Stenger et al. 1992). In Tab.2 the expected annual event rates per steradian arising from AGNs are given. Scaling to the corresponding diffuse photon fluxes leaves relatively little room for speculation regarding the energy flux in the diffuse neutrino radiation if the photon fluxes indeed result from proton-initiated cascades. The putative AGN fluxes provide the most important background for direct searches of weak-interaction decay channels of dark matter candidates.

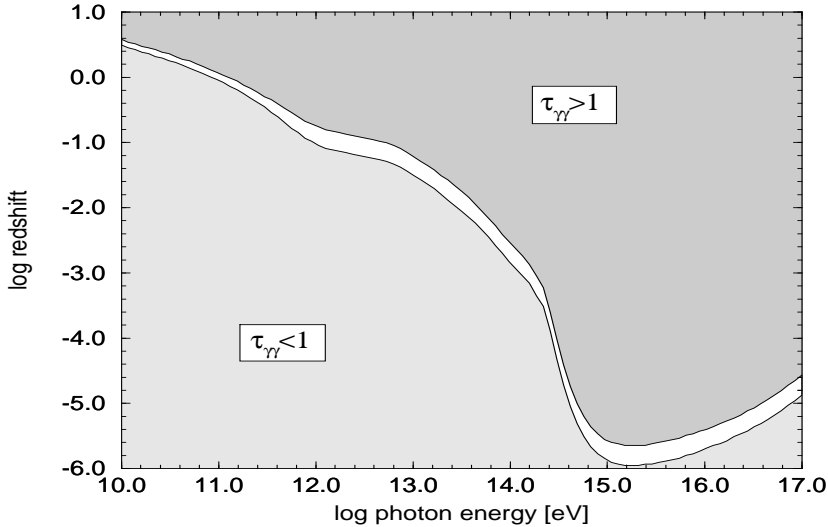

Figure 7: The gamma ray horizon due to photon-photon pair absorption for two values of the Hubble constant (*upper curve*: $H_\circ = 100$ km s$^{-1}$ Mpc$^{-1}$, *lower curve*: $H_\circ = 50$ km s$^{-1}$ Mpc$^{-1}$

# 6 The gamma ray horizon

High energy gamma ray photons from extragalactic sources must propagate through a diffuse soup of low energy photons produced by other galaxies and by the big bang before they reach the observer (Fig.8). If the gamma ray photon energy is high enough, i.e. larger than $E = 1/(\epsilon(1+z)^2/0.5 \text{ eV})$ TeV, there is a probability of $1 - \exp[-\tau_{\gamma\gamma}]$ that the photon will be absorbed by pair creation $\gamma + \gamma \to e^+ + e^-$ (cf. Sect.2). Since the pair creation optical depth rises with energy as $\tau_{\gamma\gamma} \propto E^\alpha$ for a power law distribution of target photons, cosmic absorption causes a steep exponential turnover above the energy $E^*$ where $\tau_{\gamma\gamma}(E^*) = 1$ due to *external* absorption (Fig.9). Since $\tau_{\gamma\gamma} \simeq n_\text{b}\sigma_{\gamma\gamma}D$, a measurement of $E^*$ entails information on the product $n_\text{b}D \propto n_\text{b}/H_\circ$. If the diffuse infrared background would be known from observations, one could therefore infer the Hubble constant at an unprecedented distance scale where the Hubble flow is faster than any possible mass flows due to inhomogeneous mass distributions (De Jager et al. 1994). On the other hand, since a direct measurement of $n_\text{b}$ is difficult due to zodiacal and cirrus light pollution, one can infer the photon background instead assuming a given value of the Hubble constant (Stecker et al. 1992). The photon background density $n_\text{b}$ is a strong function of galaxy evolution: Early galaxy formation (cold dark matter) implies very large values of $n_\text{b}$, whereas late galaxy formation (cold + hot dark matter) implies low values of $n_\text{b}$. For the computation of the optical

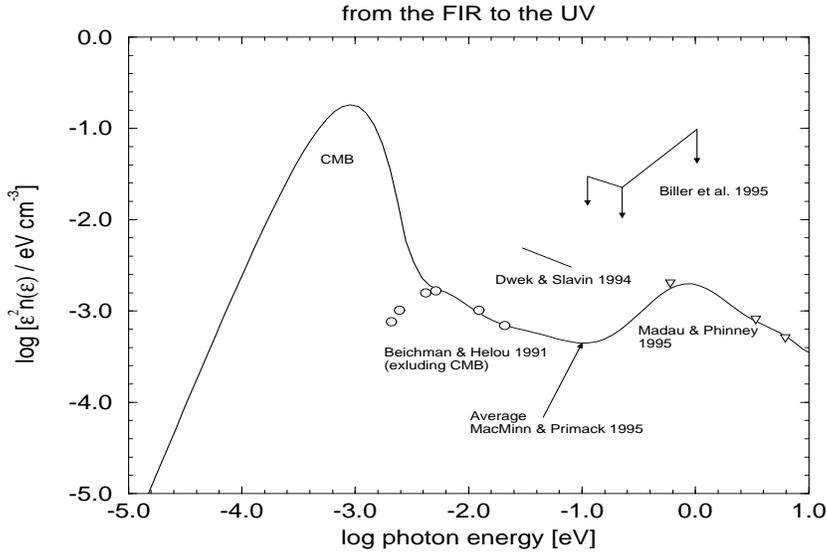

Figure 8: The cosmic background radiation from microwaves (3K radiation) over FIR photons emitted by dust to UV photons emitted by early galaxies

depth we use the geodesic radial displacement function

$$\frac{dl}{dz} = \frac{c}{H_\circ}[(1+z)E(z)]^{-1} \tag{30}$$

where $E(z)$ is given by Eq.(13.3) in Peebles (1993). For a cosmological model with $\Omega = 1$ and $\Lambda = 0$ the function $E(z)$ simplifies to $(1+z)^{3/2}$. Hence we obtain the optical depth

$$\tau_{\gamma\gamma}(E, z_\circ) = \int_0^{z_\circ} dz \frac{dl}{dz} \int_{-1}^{+1} d\mu \frac{1-\mu}{2} \int_{\epsilon_{\rm th}}^{\infty} d\epsilon n_{\rm b}(\epsilon)(1+z)^3 \sigma_{\gamma\gamma}(E, \epsilon, \mu, z)$$

$$= \frac{c}{H_\circ} \int_0^{z_\circ} dz (1+z)^{1/2} \int_0^2 dx \frac{x}{2} \int_{\epsilon_{\rm th}}^{\infty} d\epsilon n_{\rm b}(\epsilon) \sigma_{\gamma\gamma}(E, \epsilon, x-1, z) \tag{31}$$

for a non-evolving present-day background density $n_{\rm b}$, i.e. $n'_{\rm b}(z, \epsilon')d\epsilon' = (1+z)^3 n_{\rm b}(\epsilon)d\epsilon$ where the dash indicates comoving-frame quantities. The shape of the present-day diffuse background density $n_{\rm b}(\epsilon)$ is obtained by averaging over various galaxy formation models presented by MacMinn & Primack (1995). We multiplied this shape by a small factor to obtain agreement with the background density (including the contribution from the 2.7 K microwave background) estimated by Beichman & Helou (1992) in the FIR (for a modest galaxy luminosity or density evolution $\gamma = 2$), and by Madau & Phinney (1995) in the NIR through UV range (Fig.8). The predicted background densities depend sensitively on which density or luminosity evolution and maximum redshift are assumed and can vary by an order of magnitude. Hence it is very important to tighten existing constraints on the actual background

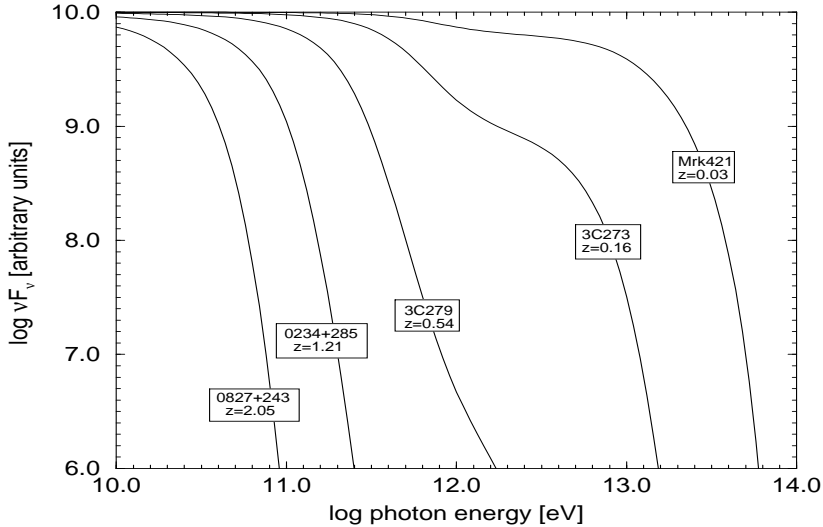

Figure 9: Templates of absorbed gamma ray spectra assuming for simplicity $\nu F_\nu = const.$ and no intrinsic cutoff ($H_\circ = 50$ km s$^{-1}$ Mpc$^{-1}$)

density (Biller *et al.* 1995) by obtaining better galaxy luminosity functions from deep galaxy surveys or by direct measurements of blazar gamma ray (external!) turnover detections.

We numerically integrate the optical depth function and solve for the gamma ray horizon $\tau_{\gamma\gamma}(E,z) = 1$. Results are shown in Fig.7 for two values of the Hubble constant. Obviously, to measure the diffuse photon background from the era of galaxy formation requires a measurement of gamma ray absorption over a wide range of gamma ray energies from 10 GeV ($z \sim 2$) to 30 TeV ($z \sim 0.01$). The large number of gamma ray emitting AGNs discovered by EGRET provide the beacons at the gamma ray horizon which can be used for this challenging program. The minimum and maximum redshifts of these blazars are $z_{\rm min} = 0.0308$ (Mrk421, seen at TeV by two independent groups) and $z_{\rm max} = 2.172$ (S5 0836+710). For the important gamma ray range from 10-100 GeV, the silicon strip detector GLAST would be of great importance. In the energy range from 0.1-10 TeV ground-based Čerenkov telescopes and in the 10-100 TeV range scintillator arrays such as HEGRA should be dedicated to measure the gamma ray horizon Fig.7 using blazars at different redshifts.

Another application of gamma ray absorption measurements is to constrain the distance of GRBs (Alexandreas et al. 1994, Hurley 1996). If the bursts are distributed to redshifts of $z_{\rm max} \sim 2$ as indicated by their apparent brightness distribution (Cohen & Piran 1994), and if they have high energy power law tails as suggested by the detection of GRBs in the field of view of EGRET (Hurley et al. 1994, Dingus 1995), the number of GRBs at a given gamma ray threshold energy $E_o$ is greatly reduced below the number visible at BATSE energies ($\sim 100$ keV). All GRBs at distances greater than the gamma ray horizon are pair-absorbed by the intervening low energy photon soup. Thus, if the curve shown in Fig.10 can be measured by gamma astronomical methods, this would render a local origin of bursts in an extended halo highly unlikely. If the number of bursts decreases more rapidly than shown in Fig.10, this would argue for cutoffs intrinsic to the burst spectra and would not allow for a conclusion regarding the distance distribution of bursts. The third possibility is that the observed number of bursts at high energies is larger than shown in Fig.10. This would unequivocally rule out cosmological models of GRBs. E.g., a single detection of a burst at scintillator energies $50 - 100$ TeV would be sufficient to rule out a cosmological origin.

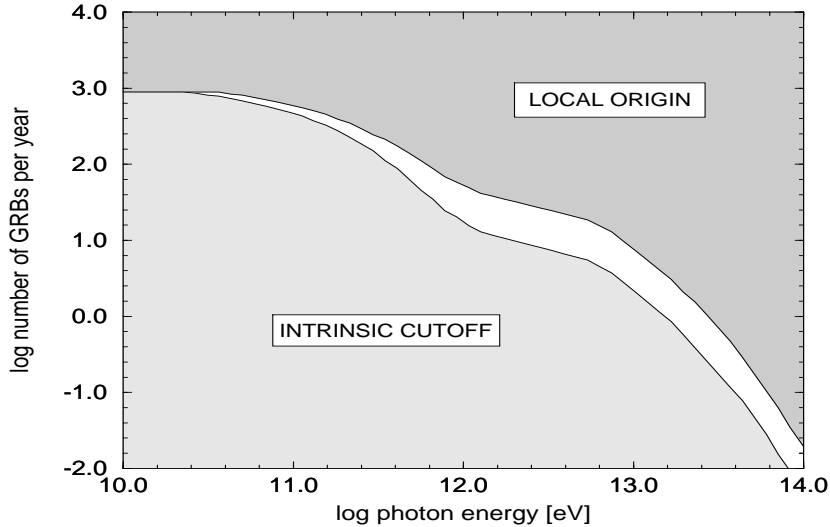

Figure 10: The number of gamma ray bursts at high photon energies. An excess of bursts over the predicted distribution would indicate a local origin in a galactic halo, a lack of bursts would indicate that intrinsic cutoffs are important. *Upper curve:* $H_o = 100$ km s$^{-1}$ Mpc$^{-1}$. *Lower curve:* $H_o = 50$ km s$^{-1}$ Mpc$^{-1}$.

# 7 Conclusions

Relativistic winds and the formation of collisionless shocks play a key role in the production of the extremely bright non-thermal gamma ray emission from many compact astrophysical objects. It may be conjectured that *relativistic* collisionless shocks are highly efficient in producing ultrarelativistic particles. A major theoretical effort to model particle acceleration at relativistic shocks is mandatory. The specific radiation mechanism responsible for the gamma rays is still controversial and depends on which particle species is accelerated to which maximum energy. One possibility is that protons reach ultrahigh-energies due to their negligible energy losses at low energies. The main proton energy loss in collisionless shocks is the photo-production of secondaries in the high density synchrotron radiation field produced by the accelerated electrons. Unsaturated synchrotron cascades induced by the protons at energies $10^8 - 10^{11}$ GeV are characterized by a generic spectrum rising in the X-ray range and reaching energies of up to $\sim 10$ TeV. The spectra match the observed broadband continua of blazars and may indicate shocks in relativistic jets. Contemporaneous multi-frequency observations of gamma ray emitting blazars from radio frequencies to TeV photon energies will improve our understanding of particle acceleration and will probe the composition of the plasma emerging from the vicinity of the putative black holes residing in AGNs.

The highest proton energies of order $\sim 3 \times 10^{11}$ GeV are most likely reached in hot spots of FR II radio galaxies implying (i) a non-thermal X-ray spectrum flatter and stronger than synchrotron-self-Compton radiation (Mannheim et al. 1991) and (ii) a flux of escaping protons which could explain the local ultrahigh-energy cosmic rays (Rachen & Biermann 1993). The recently discovered non-thermal X-rays from the hot spots of Cyg A and the indication of an excess of ultrahigh-energy cosmic rays from the supergalactic plane (Stanev et al. 1995) are in agreement with the predictions. However, a synchrotron-self-Compton origin of the X-rays is ruled out only if $B_{\rm hs} > 2 \times 10^{-4}$ G and the statistical significance of the ultrahigh-energy cosmic ray/supergalactic plane association is not yet fully compelling. An *unambiguous* signature of proton acceleration in AGNs is the emission of ultrahigh-energy neutrinos. Neutrino detection with underwater and underice detectors currently under construction is feasible. The predicted astrophysical sources of neutrinos provide the most important background for direct searches of supersymmetric dark matter candidates.

Irrespective of the nature of the gamma rays from AGNs and GRBs, their absorption by intervening diffuse radiation fields can be used as a plain probe of the cosmological distance scale or the density of the diffuse far-infrared-to-ultraviolet photons. A program to constrain models of galaxy formation requires blazar observations in the energy range 10 GeV to 10 TeV with satellite gamma ray detectors such as GLAST and ground-based detectors such as HEGRA and Whipple.